# Electron doping of proposed quantum spin liquid kagomé Zn-Cu hydroxyl-halides produces localized states in the band gap

Qihang Liu[1,2,*], Qiushi Yao[2], Z. A. Kelly[3], C. M. Pasco[3], T. M. McQueen[3,4], S. Lany[5] and Alex Zunger[1,*]

[1]*Renewable and Sustainable Energy Institute, University of Colorado, Boulder, CO 80309, USA*

[2]*Department of Physics and Shenzhen Institute for Quantum Science and Technology, Southern University of Science and Technology, Shenzhen 518055, China*

[3]*Department of Chemistry, Institute for Quantum Matter, and Department of Physics and Astronomy, The Johns Hopkins University, Baltimore, MD 21218*

[4]*Department of Materials Science and Engineering, The Johns Hopkins University, Baltimore, MD 21218*

[5]*National Renewable Energy Laboratory, Golden, Colorado 80401, USA*

[*]*E-mail:* liuqh@sustc.edu.cn; alex.zunger@colorado.edu

**Abstract**

Carrier doping of quantum spin liquids is a long-proposed route to the emergence of high-temperature superconductivity. Electrochemical intercalation in kagomé hydroxyl-halide materials shows that samples remain insulating across a wide range of electron counts. Here we demonstrate through first-principles density functional calculations corrected for self-interaction the mechanism by which electrons remain localized in various Zn-Cu hydroxyl-halides, independently of the chemical identity of the dopant – the formation of polaronic states with attendant lattice displacements and a dramatic narrowing of bandwidth upon electron addition. The same theoretical method applied to electron doping in cuprate $Nd_2CuO_4$ correctly produces a metallic state when the initially formed polaron dissolves into an extended state. Our general findings explain the insulating behavior in a wide range of "doped" quantum magnets and demonstrate that new quantum spin liquid host materials are needed to realize metallicity borne of a spin liquid.

A magnetically frustrated band insulator can form a quantum spin liquid (QSL), presenting an interacting quantum system in which spins do not order at low temperature [1]. The QSL has been theorized to offer insights into an intricate part of the mechanism for high-temperature superconductivity [2,3] if the electron count can be tuned by doping away from 1e/site while the resultant carriers become mobile. Heisenberg antiferromagnets such as synthetic Herbertsmithite or synthetic Barlowite, $Zn_xCu_{4-x}X$ with X = $[(OH)_6Cl_2]$ or $[(OH)_6BrF]$ have kagomé lattices and display several QSL fingerprints [3-6]. Interestingly, Kelly et al. recently found that insertion of as much as 0.6 Li ions (a nominal n-type dopant) per $Cu^{2+}$ into $ZnCu_3(OH)_6Cl_2$ does not show free electrons or metallic conductivity [7]. This raises the question of whether this absence of free electrons upon Li insertion into Zn-Cu based kagomé lattices could be remedied by other chemical insertion, or that Zn-Cu class kagomé structures are intrinsically impervious to addition of mobile free electrons, and thus do not offer the long-sought platform for studying high-temperature superconductivity. The latter behavior is known in insulator physics where certain compounds do not accommodate stable doping of a given polarity independent of the doping technique (e.g., n-type diamond; p-type ZnO or p-type MgO) [8,9]. Furthermore, the Fermi energy ($E_F$) in real solids can generally not be shifted at will by arbitrarily raising electron count (as attempted in Ref. [10]) because a modified trial $E_F$ can reside in different electronic states that alter the chemical bonding, so atomic displacements might ensue that redefine self-consistently a new $E_F$ leading to its pinning via a self-regulating response [11]. Therefore, theoretical work is needed to understand and predict the role of carrier insertion in kagomé QSL, particularly the interplay between local structural disorder and possible tendency toward electron localization.

Use of density-functional theory (DFT) with current exchange-correlation functionals may not be sufficient to provide an answer because of the "delocalization error" [12] whereby the convex nonlinear downward bowing of the total energy with respect to electron occupation number leads to energy gain by *spreading* the wavefunction [13,14].

In this Letter, we correct the delocalization error using a "cancellation of non-linearity" (CONL) approach, allowing for systematic inquiry into the question of whether the QSL candidates $Zn_xCu_{4-x}(OH)_6BrF$ will localize or delocalize added carriers. In general, doping can perturb the states of the pristine system, even creating entirely new states inside the band gap. As illustrated by $ZnCu_3(OH)_6BrF$ (see Fig. 1), whereas the $d^9$ electrons of $Cu^{2+}$ in the *undoped system* are spread over a broad energy range in the valence band as a result of large exchange and crystal field splitting, upon *doping an electron* there is strong localization around a single $Cu^{1+}$ site, with local atomic displacements and magnetic moments indicating a $d^9$ to $d^{10}$ transition. This establishes a detailed material-dependent theory pointing to other localization mechanisms in the behavior of QSL doping. Our predicted electron doping characteristics for Zn-Cu hydroxyl-halides leading to a polaron deep in the band gap is in sharp contrast with electron doping in the $Cu^{2+}$ cuprates (e.g., $Nd_2CuO_4$), where the same method of doping calculation shows that the polarons initially formed close to the conduction band at low electron concentration tend to dissolve into extended state through polaron overlap, leading at ~15% doping to a semiconductor–metal transition that is consistent with experiments [15-17].

*Potential bottlenecks to doping and our strategy to prove/disprove localization:* Failure to create free carriers upon insertion of a nominal dopant atom into the lattice can have many reasons, so one needs to eliminate first the trivial reasons before concluding localization. Several factors can cause no free electrons following insertion of a nominal doping agent (such as Li):

(a) The inserted atom decomposes the host system (say, forming a secondary phase with some of the host atoms, as is the case with reactive dopants). Such phase changes are known in many materials (e.g., tetragonal to hexagonal in $K_{1-x}Ni_2Se_2$ [18]), and are readily observed by normal diffraction techniques. Doping holes into herbertsmithite or barlowite has not been successful to date, instead resulting in decomposition [7].

(b) If (a) does not happen, still the inserted atom may have negligible solubility in the host system (say, due to size mismatch) so not enough dopants to create free carriers (e.g., doping Sn into GaAs).

(c) If (a) and (b) do not happen, the soluble atom may end up forming a deep (say, mid-gap) level that is not ionizable at the operation temperatures. For example, ZnO:N is deep acceptor [19].

(d) If (a)–(c) do not happen, the carriers released by the dopant can instigate the formation of an intrinsic counter defect that compensated the effect of the intentional dopant. Examples include the formation of Zn vacancies (acceptors) in response to n-type doping by Al of ZnO [20], and Ga vacancy in response to n-type doping (at the carrier concentration of $10^{19}$ cm$^{-3}$) to Ga$_2$O$_3$ [21].

(e) If (a)–(d) do not happen, uncompensated free electrons can become spatially trapped by a small polaron formation ("digging its own grave" by atomic relaxation) [22]. To remove the dependence on the chemical nature of the dopant atom we have added electrons to the system without a specific impurity, just by shifting $E_F$ up. Such so-called "non-chemical doping" (analogous to gating) is to circumvent mechanisms (a)–(d) so as to learn how the host system reacts to free electrons.

Currently available exchange-correlation functional ($E_{xc}$) in DFT usually fails systematically to predict localized polaron states even in cases where its formation is a fact [13,23-27] because the self-interaction error often leads to an unrealistic delocalized wavefunction. A good correction needs to be given to fulfill the so-called generalized Koopmans condition [13,14]

$$\Delta_{nk} = E(N-1) - E(N) + eig(N) = 0, \quad (1)$$

where $E(N-1) - E(N)$ denotes the total energy cost to remove an electron from the electron-doped system, and $eig(N)$ the single-particle energy of the highest occupied state in the electron-doped system. The cancellation of the non-linearity $\Delta_{nk} = 0$ can be achieved by the currently unknown exact $E_{xc}$ or by deliberate adjustments to the approximate $E_{xc}$. One way to remedy the non-linearity is hybrid functional [28] containing a mix of exact exchange from Hartree–Fock (HF) theory with the $E_{xc}$ from DFT. The other method is introducing a potential operator that acts only on the doping states to restore the generalized Koopmans condition (see Supplementary Materials for details of the DFT+U potential and the nonlocal external potential [29]). Here we apply both methods on kagomé Zn-Cu hydroxyl-halides for the cross-validation of electron polaron formation.

To reach the generality of the electron doping behavior of $Cu^{2+}$ hydroxyl halides we consider a variety of experimentally observed structures of $Zn_xCu_{4-x}(OH)_6BrF$. The first synthesize report suggested a $P6_3/mmc$ structure with Zn substitution widely ranging from $x = 0$ to 1 [6,42,43], while some of the authors reported an orthorhombic Cmcm structure with $x$ less than 0.5 [44]. DFT local optimization is performed on all these input structures before calculating the electronic structures.

*Results for Zn-Cu hydroxide:* As a representative, we first consider $ZnCu_3(OH)_6BrF$ with a $P6_3/mmc$ crystal structure, as shown in Fig. 1a. For simplicity we assume a ferromagnetic alignment within each Cu kagomé layer and an antiferromagnetic ordering between the neighboring Cu layers. The exchange interaction and doping effects of other magnetic configurations are summarized in the Supplementary Materials [29]. From the calculations we find that the response of the material to an added electron by localizing it does not depend on the details of the spin order. Indeed, this is in line with diverse types of materials that localize incoming electrons [45], suggesting that such description for added carriers is generally appropriate and needs not be specialized to QSL phases. A unit cell contains two Cu kagomé layers with AA stacking, i.e., 6 Cu atoms. Each Cu is coordinated by four oxygen atoms forming a rectangle network. Minimizing the total energy wrt. atomic positions (fixed cell shape) shows that in the undoped compound the inter-planar O-Cu-O angles are 85.6 and 94.4 degree, and the Cu-O bond length is 1.95 Å. The counterparts of experiment value are 84.3, 95.7 degree, and 1.94 Å, respectively [6]. Our calculations by using hybrid functionals (in HSE06 form [28]) show that $ZnCu_3(OH)_6BrF$ refers to a positive charge-transfer band insulators (with a band gap of 3.5 eV [46]), i.e., the conduction band (upper Hubbard band) is dominantly contributed by Cu-$d_{x^2-y^2}$ state, whereas the valence band represents an hybridization between Cu-d and O-p states. The projected DOS in Fig. 1b indicates a $d^9$ configuration of each $Cu^{2+}$ ion.

Doping one electron into the 144-atom supercell (24 Cu atoms per cell) followed by geometry optimization shows the optimized structure indicating the formation of $Cu^{1+} d^{10}$ self-trapped states. Such polaron formation is accompanied by elongation of the bond length of all the four Cu-O coordinations from 1.95 Å to 2.09 Å and reduction of the magnetic moment from 0.7 to 0 μB. Interestingly, compared with the undoped system in which d$^9$ electrons of Cu$^{2+}$ are spread over a broad energy range in the valence band (see Fig. 1b), this broad distribution of levels is bunched into a narrow range of highly localized d$^{10}$ states inside the band gap upon electron doping (see Fig. 1c). This fact suggests that while spreading d$^9$ states are part of the extended system with band dispersion, narrow d$^{10}$ states are in essence defect states from a single site. When adding the last electron of the d-shell of Cu, the enhanced Coulomb interaction raises the energy of all the other d-electrons, which can be in the order of several eV. In addition, in the d$^{10}$ Cu$^{1+}$ shell some spin-splitting remains, due to the asymmetry between spin-up and spin-down in the rest of the lattice. The charge density of the highest occupied polaronic state (see Fig. 1d) shows a d$_{x2-y2}$ orbital character localized on Cu$^{1+}$ ion. By adding one electron into a unit cell with 6 Cu atoms, we show that this mid-gap Cu d$^{10}$ localization is independent on different concentrations we considered [29].

While the hybrid functional calculation fulfills Koopmans condition approximately, out CONL correction to standard DFT guarantees linearity by construction. With standard DFT+U only, the final structure after relaxation ends up without the local symmetry breaking. As a result, the added electron moves $E_F$ up onto the conduction band and forms an extend state distributing throughout the whole cell, leading to a metallic feature (see Fig. 2a and 2c). As long as the parametrized strength of the onsite electron potential $\lambda_e$ is large enough (exceeds a critical value $\lambda_{cr}$ ~1.3 eV) [29], we can stabilize the structure with local symmetry breaking, and the Cu electron polaron forms, as shown in Fig. 2b and 2d. The results of Cu$^{1+}$ polaron are qualitatively similar with the calculations using hybrid functional, pertaining the main feature of the d$^9$-d$^{10}$ transition inside the gap, as well as the doped electron localized at the d$_{x2-y2}$ orbital of one Cu$^{1+}$ ion.

Figure 3a shows the evolution of structural and magnetic properties around a Cu ion in ZnCu$_3$(OH)$_6$BrF as a function of $\lambda_e$. It is apparent that during the d$^9$ – d$^{10}$ transition, the Cu-O bond length increases and the local magnetic moment quenches. An appropriate choice of $\lambda_e$ should fulfill the generalized Koopmans condition Eq. (1). Figure 3b shows the non-Koopmans energy $\Delta_{nk}$, defined as $E(N) - E(N-1) - eig(N)$, as a function of $\lambda_e$. We find that Eq. (1) is fulfilled at $\lambda_{lin}$ ~ 1.95 eV, at which point the linearity is correctly recovered. Since we have $\lambda_{lin} > \lambda_{cr}$, the polaronic state with local symmetry breaking is physically meaningful in presenting the electron doping of ZnCu$_3$(OH)$_6$BrF.

We next consider another spin-1/2 kagomé antiferromagnet, i.e., $Cu_4(OH)_6BrF$ (barlowite), which has also been proposed as QSL candidate [3,43]. The results of electron-doped $Cu_4(OH)_6BrF$ and its derivative with 25% interlayer Zn substitution $Zn_{0.25}Cu_{3.75}(OH)_6BrF$ (see Supplementary Materials for details [29]) indicate that the polaron nature is robust with the presence of small symmetry breaking distortions. Such universality suggests that the physics of electron localization might originate from the intralayer correlation between $Cu^{2+}$ sites and the particular planar $CuO_4$ coordination.

*Comparison with doping cuprates*: The kagomé hydroxides share important features in common with the T'-phase cuprates (e.g., $Nd_2CuO_4$ and $Pr_2CuO_4$) as the host of high-temperature superconductors [16], including (1) they both have planar $CuO_4$ local coordination; (2) they are positive charge-transfer insulators. However, in contrast with Kagomé, the T' phase was successfully doped with free carriers. To validate our CONL method we have applied it to electron doping of $Nd_2CuO_4$. We use $\lambda_e = 2$ eV on the Cu-d orbital [47], with two doping concentrations (6.25% and 12.5%). At the low doping concentration, we find that an electron polaron forms with energy in the upper part of the band gap, localized on the Cu site and accompanied with local lattice distortion, indicating an insulating phase. In contrast to $ZnCu_3(OH)_6BrF$, the polaron state in $Nd_2CuO_4$ has strong hybridization between the $Cu^{1+}$ ion and its O ligands, indicating a larger polaron radius, and thus higher possibility to become conductive due to inter-polaron overlap at a moderate doping concentration. When the doping concentration reaches 12.5%, all the configurations we considered become small-gap semiconductor or even band conductor with substantial polaron overlap (see Fig. 4). Our finding is in agreement with the conductivity measurement of Ce-doped $Nd_2CuO_4$ and $Pr_2CuO_4$, showing a semiconductor-metal/superconductor transition at ~14% n-type doping concentration [48,49]. Further details on doping cuprates are given in the Supplementary Materials [29]. Comparing between doped quantum spin liquid candidates and T'-$Nd_2CuO_4$ provided direct validation of our calculation method.

Our finding in T'-cuprates is in close analogous of a *sp*-dominated superconductor host $BaBiO_3$ – the hole doping with dilute concentration remains the system semiconducting due to bipolaron formation [50], while strong interaction and wavefunction overlap lead the system to be metallic and even superconducting [50,51]. If the short-range deformation potential and polaron-mediated electron-phonon coupling are considered together, pairs of two small polarons can be extended enough to overlap, which is similar to Cooper pairs [16,52]. In principle, if the solubility of the dopant allows, an insulator-metal transition is expected when the doping concentration is high enough to trigger the long-range interaction between polarons, leading to conductivity. However, topochemical synthesis of electron-doped herbertsmithite have not revealed metallicity up to 0.6 Li insertion per Cu [7]. We note that compared with T'-cuprates where the local $CuO_4$ motifs form an atomic plane (180° Cu-O-Cu bond angle), in kagomé lattice they align each other with tilting (116.5° Cu-O-Cu bond angle, see Fig. 1a), providing more flexibility for bond expansion. Thus, we suggest that uniaxial pressure along c direction (to flatten the Cu-O-Cu bond angle within a plane) might provide less flexibility to form a lattice-trapped polaron and be more effective at making a dopable QSL turn into a metal/superconductor.

*Discussion and conclusion:* The power of Kohn-Sham density functional theory is that the ground-state energy and spin densities of the real-life, interacting electrons in an external spin-dependent potential $v_\sigma(\boldsymbol{r})$ can be found from an effective one-electron Schrödinger equation in a single-determinant approach. This requires, however, that the exact exchange-correlation energy functional and its functional derivative are known. Building on the central fact that DFT is an exact formal theory for the ground-state properties for the exact exchange-correlation energy functional, there is no reason why the properties noted above couldn't be captured by DFT in principle. In fact, many previously considered "classical correlated solids" have been recently treated by DFT with results that are fully consistent with experiment. For example, the existence of gaps and doping effects in "Mott insulators" such as transition-metal oxides [53-56], or $La_2CuO_4$ [57], both thought for a long time to require heavy artillery of more advanced techniques but now appear to be described reasonably well by appropriately-executed DFT. Whereas the exact ultimate functional is still unavailable, the current work represents the state of the art in DFT doping calculations [9] by correcting the leading deficiency of previous DFT doping applications via imposing the generalized Koopmans theory. Thus, we believe that it can predict the real physics without further overestimation or underestimation of the localization tendency. In addition, we note that the state of magnetic ordering can alter the band dispersions and thereby affect the localization energy, but does not in principle affect the localization mechanism. Although beyond the scope of this work, our approach is well suited for studying the interplay between magnetic order/disorder and electron localization, while fully accounting for spatial spin correlations. The spin liquid phase is expected to enhance the self-trapping localization effect because of the global spin disorder.

To summarize, here we demonstrate the mechanism of insulating behaviors upon a wide range of electron doping in various Zn-Cu hydroxyl-halides, i.e., the Cu-O manifold has an intrinsic tendency to localize added electrons into a self-trapped polaronic state. Such an electron-localization mechanism happens even without the disorder from chemical doping, standing as an important insight independent of any experiment. The doping-induced disorder, either by the randomness of the incoming carrier or the local distortion by the chemical dopant, may help further stabilize the polaron. Therefore, it is unlikely that any $Cu^{2+}$ hydroxide with the triangular motif will support free carriers when electron doped. In contrast, for $Cu^{2+}$ cuprates $Nd_2CuO_4$ a moderate electron-doping concentration leads to extended states through polaron hopping and thus conductivity. Our findings generally explain the insulating behavior in a wide range of "doped" quantum magnets, and suggest that new candidates of quantum spin liquid are needed to realize metallicity or high-temperature superconductivity by resonating valence bonds theory.

**Acknowledgements**
Work at the University of Colorado Boulder involved all of the calculations on the Zn-Cu compounds and was supported by the U.S. Department of Energy, Office of Science, Basic Energy Sciences, Materials Sciences and Engineering Division under Grant No. DE-SC0010467. Work at the Johns Hopkins University was supported by the U.S. Department of Energy, Office of Basic Energy Sciences, Division of Material Sciences and Engineering under grant DEFG02-08ER46544 involved experimental assessment of doping. Q.L. acknowledges Dr. H. Peng and J. Mei for helpful discussions and support of the National Young 1000 Talents Plan. TMM acknowledges support of the David and Lucile Packard Foundation. S.L. was supported by the US Department of Energy, Office of Science, Office of Basic Energy Sciences, Energy Frontier Research Centers, under contract no. DE-AC36-08GO28308 to NREL, where GW calculations and CONL codes were done. This work used resources of the National Energy Research Scientific Computing Center, which is supported by the Office of Science of the U.S. Department of Energy under Contract No. DEAC02-05CH11231, and resources of high-performance computing resources sponsored by the Office of Energy Efficiency and Renewable Energy.

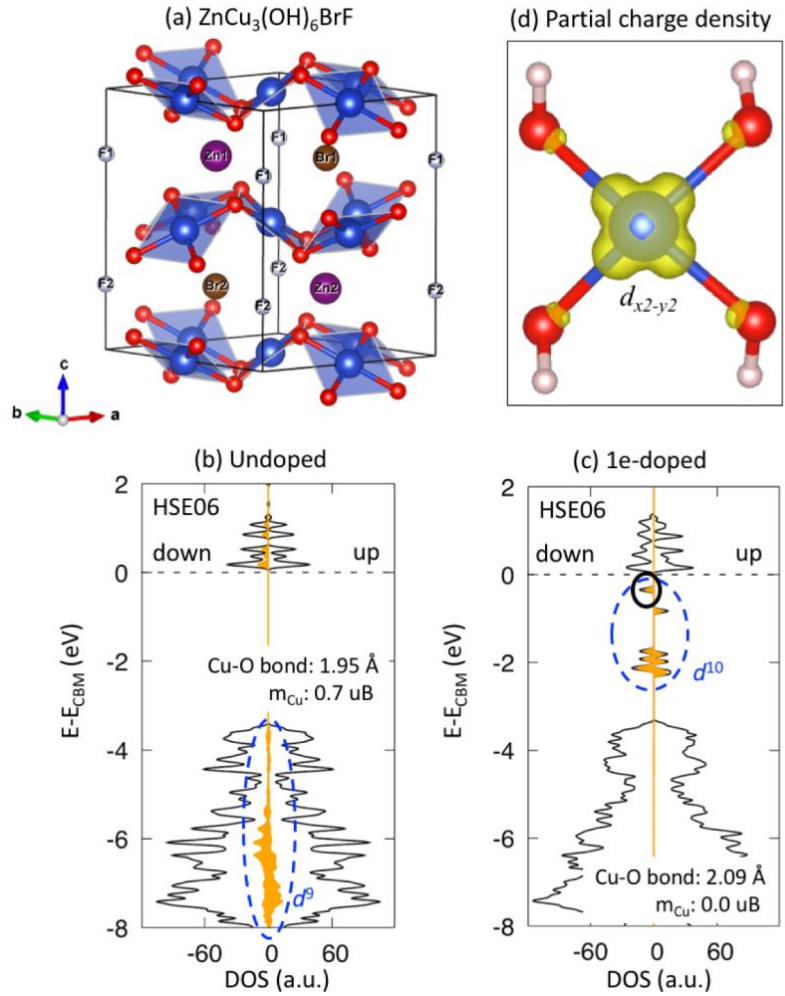

Fig. 1: (a) Perspective view of the observed kagomé crystal structure of $P6_3/mmc$ $ZnCu_3(OH)_6BrF$. Cu and O atoms are indicated by blue and red balls, respectively, while H atoms are not shown for clarity. (b,c) HSE06 calculated total density of states (DOS, black) for (b) undoped $ZnCu_3(OH)_6BrF$ and (c) non-chemical doping of 1 electron into the 144-atom supercell. The orange curves denote the projected DOS of (b) one of the $Cu^{2+}$ ion and (c) the resultant $Cu^{1+}$ polaron by doping. (d) The isosurface of charge density (yellow) of the highest occupied state (black circle in panel c).

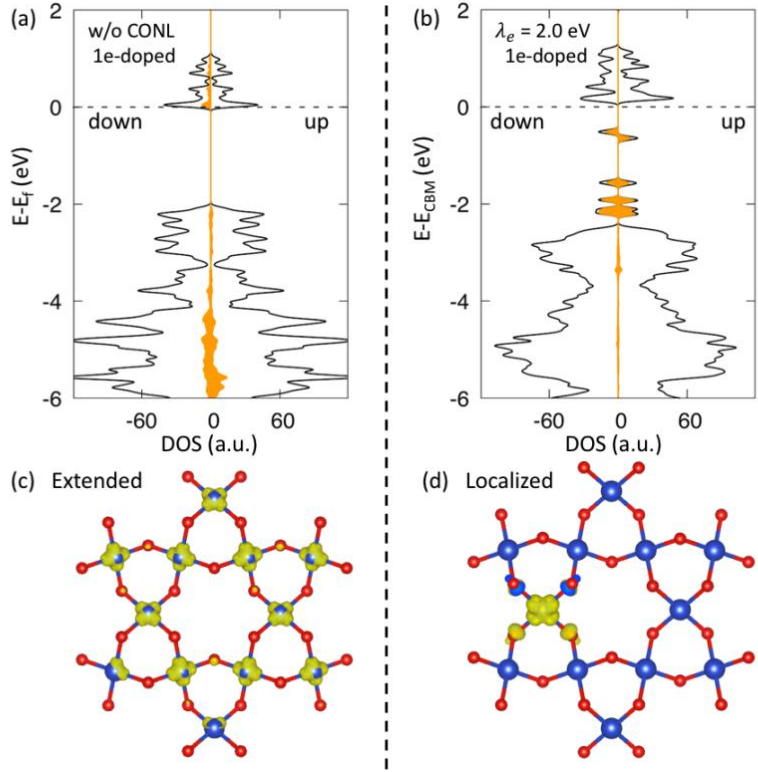

Fig. 2: Calculated DOS (black) and projected DOS of $Cu^{1+}$ ion (orange) of $ZnCu_3(OH)_6BrF$ for doping 1 electron non-chemically (a) without CONL correction and (b) with onsite potential $\lambda_e = 2$ eV. Charge density of the highest occupied states for electron doping (c) without CONL correction and (d) with onsite potential $\lambda_e = 2$ eV.

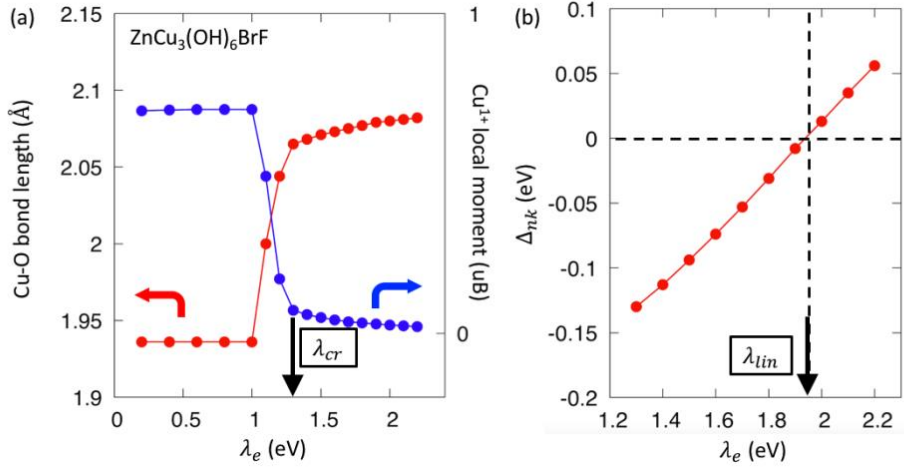

Fig. 3: (a) Calculated Cu-O bond length and magnetic moment of Cu for electron-doped ZnCu$_3$(OH)$_6$BrF, as a function of the electron-state potential strength $\lambda_e$. (b) Non-Koopmans energy $\Delta_{nk}$ as a function of the electron-state potential strength $\lambda_e$.

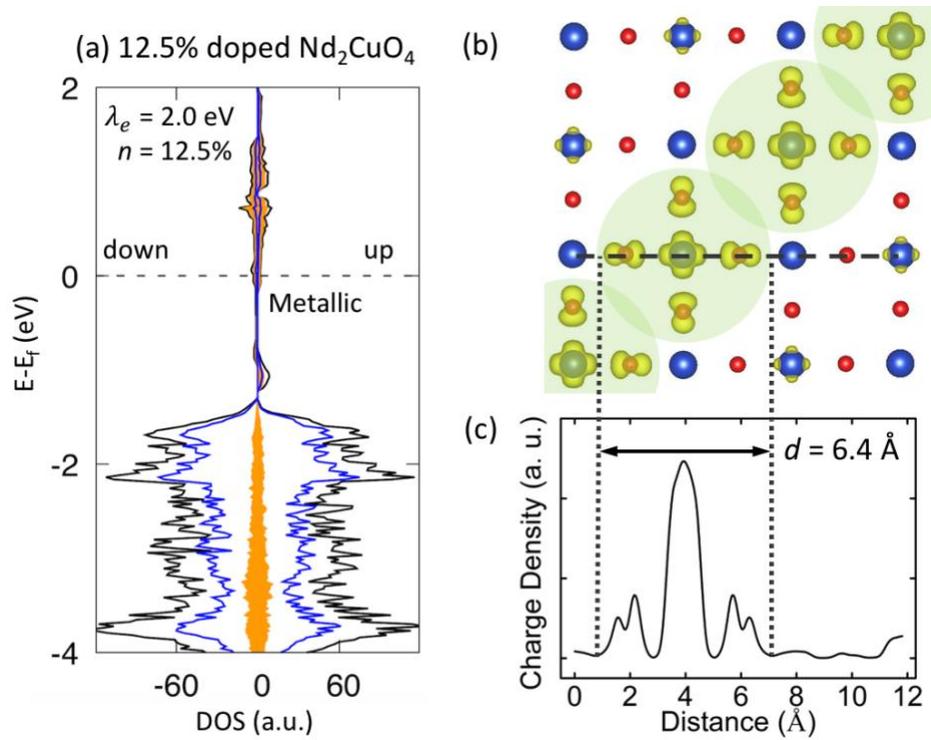

Fig. 4: (a) Calculated DOS (black) and projected DOS of Cu and O (orange and blue, respectively) of Nd$_2$CuO$_4$ for 12.5% electron doping concentration with $\lambda_e = 2$ eV. (b) Charge density of the highest occupied states (below $E_F$) shows polaron overlapping (green circles). (c) Charge density along a specific line (denoted by the dash line in panel b). The dotted lines indicate the minima of the 1D charge density, corresponding to a polaron diameter of 6.4 Å.